%% file: manuscript.tex
\documentclass[sigconf,screen]{acmart}
\def\BibTeX{{\rm B\kern-.05em{\sc i\kern-.025em b}\kern-.08emT\kern-.1667em\lower.7ex\hbox{E}\kern-.125emX}}
\usepackage{wrapfig}
\usepackage{nicefrac}
\usepackage{booktabs} 
\usepackage{pbox}
\usepackage{csvsimple}
\usepackage{amssymb}
\usepackage{mathtools}
\usepackage{rotating}
\usepackage{multicol}
\usepackage{multirow}
\usepackage{xcolor}
\usepackage[export]{adjustbox}
\usepackage{array}
\usepackage{tabularx}
\usepackage{amsthm}
\usepackage{bigstrut}
\usepackage{subcaption}
\usepackage{graphicx}
\usepackage{pgfplots}
\usepackage{soul}
\usepackage{paralist}
\usepackage{enumitem}
\usepackage{amsmath}
\usepackage{relsize}
\usepackage{parcolumns}
\usepackage{pgfplots}
\usepackage[breakable, theorems, skins]{tcolorbox}
\usepackage{marvosym}
\usepackage[flushleft]{threeparttable}
\pgfplotsset{compat=1.3}
\usepackage{color, colortbl}
\usepackage[utf8]{inputenc}
\usepackage[english]{babel}
\usepackage{tablefootnote}
\newcolumntype{M}{@{}>{\columncolor{white}[0pt][0pt]}c@{}}

\usepackage{pgf}
\usepackage{balance}
\usetikzlibrary{positioning}
\usepackage[linesnumbered,ruled,vlined]{algorithm2e}
\usetikzlibrary{arrows}
\usepackage[flushleft]{threeparttable} 

\usepackage{subcaption}
\usepackage{cleveref}

\usepackage{threeparttable}
\graphicspath{ {figures/} }

\input{spacing.tex}
\input{macros.tex}

\input{pygments.tex}

\usepackage{boldline}

\usepackage{caption}
\usepackage{pifont}

\newcommand{\eq}[1]{Eq.~\ref{eq:#1}}
\newcommand{\fig}[1]{Fig.~\ref{fig:#1}}

\newcommand{\tab}[1]{Table~\ref{tab:#1}}
\newcommand{\tion}[1]{\S\ref{sect:#1}}

\newcommand{\etal}{\hbox{\emph{et al.}}\xspace}
\newcommand{\eg}{\hbox{\emph{e.g.}}\xspace}
\newcommand{\ie}{\hbox{\emph{i.e.}}\xspace}
\newcommand{\wrt}{\hbox{\emph{w.r.t.}}\xspace}

\newcommand{\tool}{\textsc{MTFuzz}\xspace}

\newcommand{\ec}{{edge coverage}\xspace}
\newcommand{\ctx}{{context-sensitive edge coverage}\xspace}
\newcommand{\softlabel}{{approach-sensitive edge coverage}\xspace}
\newcommand{\cfg}{control-flow graph\xspace}
\newcommand{\afl}{AFL\xspace}
\newcommand{\aflfast}{AFLFast\xspace}

\newcommand{\neuzz}{\textsc{Neuzz}\xspace}

\newcommand{\squishlist}{
 \begin{list}{$\circ$}
 { 
  \setlength{\itemsep}{1pt}
   \setlength{\parsep}{1pt}
   \setlength{\topsep}{1pt}
   \setlength{\partopsep}{1pt}
   \setlength{\leftmargin}{1.5em}
   \setlength{\labelwidth}{1em}
   \setlength{\labelsep}{0.5em} } }

\newcommand{\squishlisttwo}{
 \begin{list}{$\bullet$}
 { \setlength{\itemsep}{0pt}
  \setlength{\parsep}{0pt}
  \setlength{\topsep}{0pt}
  \setlength{\partopsep}{0pt}
  \setlength{\leftmargin}{0em}
  \setlength{\labelwidth}{0.5em}
  \setlength{\labelsep}{0em} } }

\newcommand{\squishend}{
 \end{list} }

\usepackage{etoolbox,refcount}

\newcounter{countitems}
\newcounter{nextitemizecount}
\newcommand{\setupcountitems}{%
  \stepcounter{nextitemizecount}%
  \setcounter{countitems}{0}%
  \preto\item{\stepcounter{countitems}}%
}
\makeatletter
\newcommand{\computecountitems}{%
  \edef\@currentlabel{\number\c@countitems}%
  \label{countitems@\number\numexpr\value{nextitemizecount}-1\relax}%
}
\newcommand{\nextitemizecount}{%
  \getrefnumber{countitems@\number\c@nextitemizecount}%
}
\newcommand{\previtemizecount}{%
  \getrefnumber{countitems@\number\numexpr\value{nextitemizecount}-1\relax}%
}
\makeatother    
{\end{itemize}%
\unskip\computecountitems\ifnumcomp{\previtemizecount}{>}{3}{\end{multicols}}{}}

\copyrightyear{2020} 
\acmYear{2020} 
\setcopyright{acmcopyright}\acmConference[ESEC/FSE '20]{Proceedings of the 28th ACM Joint European Software Engineering Conference and Symposium on the Foundations of Software Engineering}{November 8--13, 2020}{Virtual Event, USA}
\acmBooktitle{Proceedings of the 28th ACM Joint European Software Engineering Conference and Symposium on the Foundations of Software Engineering (ESEC/FSE '20), November 8--13, 2020, Virtual Event, USA}
\acmPrice{15.00}
\acmDOI{10.1145/3368089.3409723}
\acmISBN{978-1-4503-7043-1/20/11}

\begin{document}


\title[\textsc{MTFuzz}: Fuzzing with a Multi-task Neural Network ]
{\textsc{MTFuzz}: Fuzzing with a Multi-task Neural Network}

\author{Dongdong She}
\affiliation{%
  \institution{Columbia University}
  \city{New York}
  \country{USA}}
\email{ds3619@columbia.edu}

\author{Rahul Krishna}
\affiliation{%
  \institution{Columbia University}
  \city{New York}
  \country{USA}}
\email{rk3080@columbia.edu}

\author{Lu Yan}
\affiliation{%
  \institution{Shanghai Jiao Tong University}
  \city{Shanghai}
  \country{China}}
\email{jiaodayanlu@sjtu.edu.cn}

\author{Suman Jana}
\affiliation{%
  \institution{Columbia University}
  \city{New York}
  \country{USA}}
\email{suman@cs.columbia.edu}

\author{Baishakhi Ray}
\affiliation{%
  \institution{Columbia University}
  \city{New York}
  \country{USA}}
\email{rayb@cs.columbia.edu}


\renewcommand{\shortauthors}{She, D., Krishna, R., Yan, L., Jana, S., and Ray, B.}
\input{body/0_abstract.tex}

\maketitle

\input{body/1_introduction.tex}


\input{body/2_formalization_v2}


\input{body/3_methodology.tex}

\input{body/4_implementation.tex}\input{body/6_experiments.tex}

\input{body/7_discussions}

\input{body/8_related.tex}



\input{body/9_conclusion.tex}


\balance
\bibliographystyle{ACM-Reference-Format}
\bibliography{references}
\end{document}

%% file: macros.tex
\newcommand{\bi}{\begin{itemize}}
\newcommand{\ei}{\end{itemize}}
\newcommand{\beq}{\begin{equation}}
\newcommand{\eeq}{\end{equation}}
\newcommand{\be}{\begin{enumerate}}
\newcommand{\ee}{\end{enumerate}}

\newcommand{\MTNN}{Multi-task Neural Network}

\newcommand{\inlinemath}[1]{$#1$}

\newcommand{\inlinemathit}[1]{$\mathit{#1}$}

\newcommand{\inlinemathcal}[1]{$\mathcal{#1}$}

\newcommand{\sota}{state-of-the-art\xspace}

\newcommand{\edgeids}{{edge\_id}s\xspace}

\newcommand{\xdownarrow}[1]{%
  {\left\downarrow\vbox to #1{}\right.\kern-\nulldelimiterspace}
}

\usepackage[framemethod=tikz]{mdframed}
\usetikzlibrary{shadows}
\usepackage{graphics}
\newmdenv[
    tikzsetting= {fill=blueish!15},
    skipabove=0.33em,
    skipbelow=0.33em,
    linewidth=1pt,
    innerleftmargin=4pt,
    innerrightmargin=4pt,
    innertopmargin=2pt,
    innerbottommargin=2pt,
    linecolor=gray95,
    roundcorner=2pt, 
    shadow=true,
    shadowsize=4pt,
    shadowcolor=black
]{myshadowbox}
\usepackage{tikz}

\newcolumntype{s}{>{\centering \arraybackslash \hsize=.5\hsize}X}  %

\newenvironment{result}
{\begin{myshadowbox}}
{\end{myshadowbox}}

\newcommand{\rqone}{\noindent\textsf{\textbf{RQ1: Performance.} {\normalfont How does \tool perform in comparison with other \sota fuzzers?}}\xspace}

\newcommand{\rqtwo}{\noindent\textsf{\textbf{RQ2: Contributions of Auxiliary Tasks.} {\normalfont How much does each auxiliary task contribute to the overall performance of \tool?}}\xspace}

\newcommand{\rqthree}{\noindent\textsf{\textbf{RQ3: Impact of Design Choices.} {\normalfont How do various design choices affect the performance of \tool?}}\xspace}

\newcommand{\rqfour}{\noindent\textsf{\textbf{RQ4: Transferrability.} {\normalfont How transferable is \tool?}}\xspace}

%% file: pygments.tex
\sethlcolor{lightgray}
\definecolor{pink}{RGB}{252,145,149}
\definecolor{lightpink}{RGB}{252,145,149}
\definecolor{lightgray}{gray}{0.8}
\definecolor{darkgray}{gray}{0.6}
\definecolor{Gray}{rgb}{0.88,1,1}
\definecolor{Gray}{gray}{0.85}
\definecolor{Blue}{RGB}{0,29,193}
\definecolor{MyDarkBlue}{rgb}{0,0.08,0.45} 
\definecolor{pink}{RGB}{231,95,110}
\definecolor{greenish}{RGB}{182, 231, 142}
\definecolor{orangish}{RGB}{255, 206, 144}
\definecolor{lavender}{RGB}{225, 213, 231}
\definecolor{lightergray}{rgb}{0.85, 0.85, 0.85}
\definecolor{lightestgray}{rgb}{0.95, 0.95, 0.95}
\definecolor{codebg}{HTML}{F4F4F4}
\definecolor{blueish}{RGB}{177, 206, 232}
\definecolor{gray05}{gray}{0.95}
\definecolor{gray10}{gray}{0.90}
\definecolor{gray15}{gray}{0.85}
\definecolor{gray20}{gray}{0.80}
\definecolor{gray25}{gray}{0.75}
\definecolor{gray30}{gray}{0.70}
\definecolor{gray35}{gray}{0.65}
\definecolor{gray40}{gray}{0.60}
\definecolor{gray45}{gray}{0.55}
\definecolor{gray50}{gray}{0.50}
\definecolor{gray55}{gray}{0.45}
\definecolor{gray60}{gray}{0.40}
\definecolor{gray65}{gray}{0.35}
\definecolor{gray70}{gray}{0.30}
\definecolor{gray75}{gray}{0.25}
\definecolor{gray80}{gray}{0.20}
\definecolor{gray85}{gray}{0.15}
\definecolor{gray90}{gray}{0.10}
\definecolor{gray95}{gray}{0.05}

\newcommand\Red[1]{\textcolor[rgb]{1.00,0.00,0.00}{\textbf{#1}}}

%% file: body/0_abstract.tex
\begin{abstract}
Fuzzing is a widely used technique for detecting software bugs and vulnerabilities. Most popular fuzzers generate new inputs using an evolutionary search to maximize code coverage. Essentially, these fuzzers start with a set of seed inputs, mutate them to generate new inputs, and identify the promising inputs using an evolutionary fitness function for further mutation. 

Despite their success, evolutionary fuzzers tend to get stuck in long sequences of unproductive mutations. In recent years, machine learning (ML) based mutation strategies have reported promising results. However, the existing ML-based fuzzers are limited by the lack of quality and diversity of the training data. As the input space of the target programs is high dimensional and sparse, it is prohibitively expensive to collect many diverse samples demonstrating successful and unsuccessful mutations to train the model.

In this paper, we address these issues by using a Multi-Task Neural Network that can learn a compact embedding of the input space based on diverse training samples for multiple related tasks (i.e., predicting for different types of coverage). 
The compact embedding can guide the mutation process by focusing most of the mutations on the parts of the embedding where the gradient is high. \tool uncovers $11$ previously unseen bugs and achieves an average of $2\times$ more edge coverage compared with 5 state-of-the-art fuzzer on 10 real-world programs.
\end{abstract}

\ccsdesc[500]{Software and its engineering~Software testing and debugging}

\keywords{Graybox Fuzzing, Multi-task Neural Networks, Gradient-guided Optimization, Transfer Learning}

%% file: body/1_introduction.tex
\section{Introduction}

Coverage-guided graybox fuzzing is a widely used technique for detecting bugs and security vulnerabilities in real-world software~\cite{zalewski2017american, cgc, zalewski2017american, manes2018fuzzing,wang2019superion,Nilizadeh2019,You2019SLFFW,lemieux2017fairfuzz,she2018neuzz, godefroid2008, arya2012fuzzing, evans2011fuzzing, moroz2016guided}. The key idea behind a fuzzer is to execute the target program on a large number of automatically generated test inputs and monitor the corresponding executions for buggy behaviors. However, as the input spaces of real-world programs are typically very large, unguided test input generation is not effective at finding bugs.
Therefore, most popular graybox fuzzers use evolutionary search to generate new inputs; they mutate a set of seed inputs and retain only the most promising inputs (\ie, inputs exercising new program behavior) for further mutations~\cite{zalewski2017american, wang2019superion,Nilizadeh2019,You2019SLFFW, wei2018singularity, steelix, godefroid2017learn,lemieux2017fairfuzz,jiang2018contractfuzzer}.

However, the effectiveness of traditional evolutionary fuzzers tends to decrease significantly over fuzzing time. They often get stuck in long sequences of unfruitful mutations, failing to generate inputs that explore new regions of the target program~\cite{chen2018angora, she2018neuzz, she2019neutaint}. Several researchers have worked on designing different mutation strategies based on various program behaviors (e.g., focusing on rare branches, call context, etc.)~\cite{lemieux2017fairfuzz,chen2018angora}. However, program behavior changes drastically, not only across different programs but also across different parts of the same program. Thus, finding a generic robust mutation strategy still remains an important open problem. 

Recently,  Machine Learning (ML) techniques have shown initial promise to guide the mutations~\cite{Saavedra2019, she2018neuzz, Rajpal2017}. 
These fuzzers typically use existing test inputs to train ML models and learn to identify promising mutation regions that improve coverage~\cite{she2018neuzz, godefroid2017learn, Rajpal2017, Saavedra2019}. Like any other supervised learning technique, the success of these models relies heavily on the number and diversity of training samples. 
However, collecting such training data for fuzzing that can demonstrate successful/unsuccessful mutations is prohibitively expensive due to two main reasons. First, successful mutations that increase coverage are often limited to very few, sparsely distributed input bytes, commonly known as hot-bytes, in a high-dimensional input space. Without knowing the distribution of hot-bytes, it is extremely hard to generate successful mutations over the sparse, high-dimensional input space~\cite{she2018neuzz,Saavedra2019}. 
Second, the training data must be diverse enough to expose the model to various program behaviors that lead to successful/unsuccessful mutations\textemdash this is also challenging as one would require a large number of test cases exploring different program semantics. 
Thus, the ML-based fuzzers suffer from both {\em sparsity} and {\em lack of diversity} of the target domain. 

\input{tex/fig-1.tex}

In this paper, we address these problems using Multi-Task Learning, a popular learning paradigm used in domains like computer vision  to effectively learn common features shared across related tasks from limited training data. In this framework, different participating tasks allow an ML model to effectively learn a compact and more generalized feature representation while ignoring task-specific noises. To jointly learn a compact embedding of the inputs, in our setting, we use different tasks for predicting the relationship between program inputs and different aspects of fuzzing-related program behavior (e.g., different types of edge coverage). Such an architecture addresses both the data sparsity and lack of diversity problem. The model can simultaneously learn from diverse program behaviors from different tasks as well as focus on learning the important features (hot bytes in our case) across all tasks. Each participating task will provide separate pieces of evidence for the relevance or irrelevance of the input features~\cite{ruder2017overview}.

To this end, we design, implement, and evaluate \tool,  a Multi-task Neural Network (MTNN) based fuzzing framework. Given the same set of test inputs, \tool learns to predict three different code coverage measures showing various aspects of dynamic program behavior:
\begin{enumerate}[leftmargin=*,topsep=0pt]
    \item  edge coverage:  which edges are explored by a test input~\cite{zalewski2017american,she2018neuzz}? 
\item  \softlabel: if an edge is not explored, how far off it is (i.e., approach level) from getting triggered~\cite{McMinn2004, arcuri2010does, mcminn2011search, pachauri2013automated}? 
\item  \ctx: from which call context an explored edge is called~\cite{chen2018angora,Wang2019}?
\end{enumerate}
Note that our primary task, like most popular fuzzers, is to increase edge coverage. However, the use of call context and approach level provides additional information to boost edge coverage. 

Architecturally, the underlying MTNN contains a group of hidden layers shared across the participating tasks, while still maintaining task-specific output layers. The last shared layer learns a \textit{compact embedding} of the input space as shown in Figure~\ref{fig:intro_fig}. Such an embedding captures a generic compressed representation of the inputs while preserving the important features, \ie, hot-byte distribution. We compute a saliency score~\cite{she2019neutaint} of each input byte by computing the gradients of the embedded representation \wrt the input bytes. Saliency scores are often used in computer vision models to identify the important features by analyzing the importance of that feature \wrt an embedded layer~\cite{simonyan2013deep}. By contrast, in this paper, we use such saliency scores to guide the mutation process\textemdash focus the mutations on bytes with high saliency scores. 

Our MTNN architecture also allows the compact embedding layer, once trained, to be transferred across different programs that operate on similar input formats. For example, compact-embedding learned with \tool for one \texttt{xml} parser may be transferred to other \texttt{xml} parsers. Our results (in RQ4) show that such transfer is quite effective and it reduces the cost to generate high quality data from scratch on new programs which can be quite expensive. 
Our tool is available at~\href{https://git.io/JUWkj}{https://git.io/JUWkj} and the artifacts available at \href{https://doi.org/10.5281/zenodo.3903818}{doi.org/10.5281/zenodo.3903818}.

We evaluate \tool on 10 real world programs against 5  state-of-the-art fuzzers. \tool covers at least 1000 more edges on 5 programs and several 100 more on the rest. \tool also finds a total of 71 real-world bugs (11 previously unseen) (see RQ1). When compared to learning each task individually, \tool offers significantly more edge coverage (see RQ2). Lastly, our results from transfer learning show that the compact-embedding of \tool can be transferred across parsers for xml and elf binaries.

Overall, our paper makes the following key contributions:

\begin{itemize}[leftmargin=*]
\item
We present a novel fuzzing framework based on multi-task neural networks called \tool that learns a compact embedding of otherwise sparse and high-dimensional program input spaces. Once trained, we use the salience score of the embedding layer outputs \wrt the input bytes to guide the mutation process.

\item
Our empirical results demonstrate that \tool is significantly more effective than current state-of-the-art fuzzers. On $10$ real world programs, \tool achieves an average of $2\times$ and up to $3\times$ edge coverage compared to Neuzz, the state-of-the-art ML-based fuzzer. \tool also finds $11$ previously unknown bugs other fuzzers fail to find. The bugs have been reported to the developers.

\item
We demonstrate that transferring of the compact embedding across programs with similar input formats can significantly increase the fuzzing efficiency, \eg, transferred embeddings for different file formats like ELF and XML can help \tool to achieve up to $14\times$ edge coverage compared to state-of-the-art fuzzers.

\end{itemize}


%% file: tex/fig-1.tex
\begin{figure*}[tbp!]
    \centering
    \includegraphics[width=0.9\linewidth]{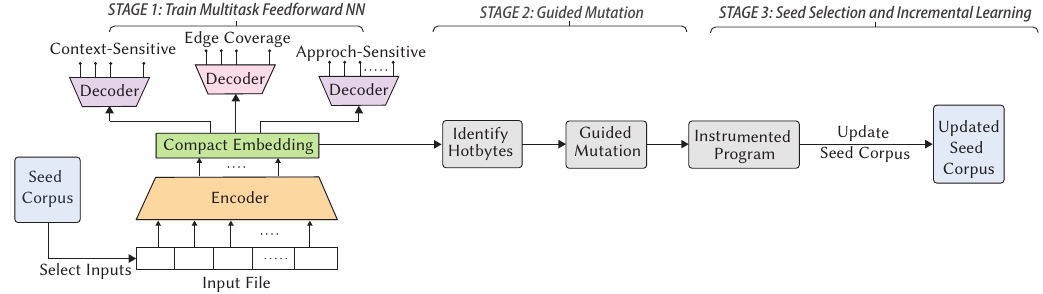}
    \caption{\small Overview of \tool}
    \label{fig:intro_fig}
\end{figure*}

%% file: body/2_formalization_v2.tex
\section{Background: Multi-task Networks}
\label{sect:theory}
\label{sect:multitask_learning}
Multi-task Neural Networks (MTNN) are becoming increasingly popular in many different domains including optimization~\cite{argyriou2008convex, gong2014efficient}, natural language processing~\cite{collobert2008unified, Binge2017}, and computer vision~\cite{standley2019tasks}. 
The key intuition behind MTNN is that it is useful for related tasks to be learned jointly so that each task can benefit from the relevant information available in other tasks~\cite{caruana1996algorithms,caruana1997multitask,standley2019tasks,zamir2018taskonomy}.
For example, if we learn to ride a unicycle, a bicycle, and a tricycle simultaneously, experiences gathered from one usually help us to learn the other tasks better~\cite{zhang2017survey}. 
In this paper, we use a popular MTNN architecture called hard parameter sharing~\cite{caruana1997multitask}, which contain two groups of layers (see \fig{bg_fig_1}): a set of initial layers shared among all the tasks, and several individual task-specific  output layers. 
The shared layers enable a MTNN to find a \textit{common feature representation} across all the tasks. The task specific layers use the shared feature representation to generate predictions for the individual tasks
~\cite{kokkinos2017ubernet,standley2019tasks,ruder2017overview}. 

\noindent\textbf{MTNN Training.} While MTNNs can be used in many different ML paradigms,  in this paper we primarily focus on supervised learning. We assume that the training process has access to a training dataset $\mathcal{X}=\{x_1, x_2, ..., x_n\}$. The training data contains the ground truth output labels for each task. We train the MTNN on the training data using standard back-propagation to minimize a multi-task loss.

\noindent\textbf{Multi-task Loss.}~An MTNN is trained using a multi-task loss function, $\mathcal{L}$. We assume that each individual task $\tau_i$ in the set of tasks $\mathcal{T}=\{\tau_1, \tau_2, ..., \tau_m\}$ has a corresponding loss function $\mathcal{L}_i$. The multi-task loss is computed as a weighted sum of each individual task loss. More formally, it is given by $\mathcal{L}=\sum^{m}_{i=1}\alpha_i\cdot\mathcal{L}_i$. Here, $\alpha_i$ represents the weight assigned to task $i$. The goal of training is to reduce the overall loss. In practice, the actual values of the weights are decided based on the relative importance of each task. Most existing works assign equal weights to the tasks~\cite{uhrig2016pixel, teichmann2018multinet, liao2016understand}. 

The multi-task loss function forces the shared layer to learn a general input representation for all tasks offering two benefits:

\noindent\textit{1) ~Increased generalizibility}. The overall risk of overfitting in multi-task models is reduced by an order of $m$ (where $\mathit{m}$ is the number of tasks) compared to single task models~\cite{baxter1997bayesian}. Intuitively, the more tasks an MTNN learns from, the more general the compact representation is in capturing features of all the tasks. This prevents the representation from overfitting to the task-specific features.

\noindent\textit{2) ~Reduced sparsity}. The shared embedding layer in an MTNN can be designed to increase the compactness of the learned input representation. Compared with original input layer, a shared embedding layer can achieve same expressiveness on a given set of tasks while requiring far fewer nodes. In such compact embedding, the important features across different tasks will be boosted with each task contributing its own set of relevant features~\cite{ruder2017overview}.

%% file: body/3_methodology.tex
\section{Methodology}
\label{sect:our_method}
This section presents a brief overview of \tool that aims to maximize \ec with the aid of two additional coverage measures: \ctx and \softlabel using multi task learning.~\Cref{fig:intro_fig} illustrates an end-to-end workflow of the proposed approach. The first stage trains an MTNN to produce a compact embedding of an otherwise sparse input space while preserving information about the hot bytes \ie, the input bytes have the highest likelihood to impact code coverage~(\Cref{sect:stage-1}). 
The second stage identifies these hot bytes and focuses on mutating them~(\Cref{sect:stage-2}). Finally, in the third stage, the seed corpus is updated with the mutated inputs and retains only the most interesting new inputs~(\Cref{sect:stage-3}).

\subsection{Modeling Coverage as Multiple Tasks}
\label{sect:tasks}

The goal of any ML-based fuzzers, including \tool, is to learn a mapping between input space and code coverage. The most common coverage explored in the literature is \ec, which is an effective measure and quite easy to instrument. However, it is coarse-grained and misses many interesting program behavior (\eg, explored call context) that are known to be important to fuzzing. One workaround is to model path coverage by tracking the program execution path per input. However, keeping track of all the explored paths can be computationally intractable since it can quickly lead to a state-space explosion on large programs~\cite{Wang2019}. As an alternative, in this work, we propose a middle ground: we model the \ec as the primary task of the MTNN, while choosing two other fine-granular coverage metrics (\softlabel and \ctx) as auxiliary tasks to provide useful additional context to \ec. 

\subsubsection{Edge coverage:~Primary Task.~}
\label{sect:ec}

Edge coverage measures how many unique control-flow edges are triggered by a test input as it interacts with the program. 
It has become the de-facto code coverage metric~\cite{zalewski2017american,she2018neuzz,lemieux2017fairfuzz,You2019SLFFW} for fuzzing. 
We model edge coverage prediction as the {\em primary task} of our multi-task network, which takes a binary test case as input and predicts the edges that could be covered by the test case. 
For each input, we represent the edge coverage as an \textit{edge bitmap}, where value per edge is set to \texttt{1} or \texttt{0} depending on whether the edge is exercised by the input or not. 

In particular, in the control-flow-graph of a program, an edge connects two basic blocks (denoted by \texttt{prev\_block} and \texttt{cur\_block})~\cite{zalewski2017american}. A unique $edge\_id$ is obtained as: $hash(\texttt{prev\_block}, \texttt{cur\_block})$.  
For each $edge\_id$, there is a bit allocated in the bitmap. For every input, the \edgeids in the corresponding edge bitmap are set to \texttt{1} or \texttt{0}, depending on whether or not those edges were triggered.

\subsubsection{Approach-Sensitive Edge Coverage:~Auxiliary Task 1.~}
\label{sect:softlabel}
\begin{figure}[h] 
    \centering
    \includegraphics[width=0.4\linewidth]{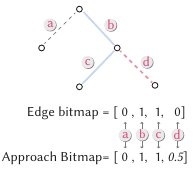}
    \caption{\small Approach Bitmap vs. Edge Bitmap. The edge `d' has a visited parent edge `b' and is thus marked 0.5 in the approach bitmap.}
    \label{fig:sl_eg}
\end{figure}
For an edge that is not exercised by an input, we measure how far off the edge is from getting triggered. Such a measure provides additional contextual information of an edge. For example, if two test inputs failed to trigger an edge, however one input reached ``closer'' to the unexplored edge than the other,  traditional  edge coverage would treat both inputs the same. However, using a proximity measure, we can discern between the two inputs and mutate the closer input so that it can reach the unexplored edge. 
To achieve this, \softlabel extends \ec by offering a distance measure that computes the distance between an unreached edge and the nearest edge triggered by an input. This is a popular measure in the search-based software engineering literature~\cite{McMinn2004,mcminn2011search,arcuri2010does}, where  instead of assigning a binary value (0 or 1), as in edge bitmap, \textit{approach level} assigns a numeric value between 0 and 1 to represent the edges~\cite{Phil2004Hyber}; if an edge is triggered, it is assigned \texttt{1}. However, if the edge is \textit{not} triggered, but one of its parents are triggered, then the non-triggered edge is assigned a value of $\beta$ (we use $\beta=0.5$). If neither the edge nor any of its parents are triggered, it is assigned $0$. This is illustrated in \fig{sl_eg}. 
Note that, for a given edge, we refrain from using additional ancestors farther up the \cfg to limit the computational burden. The approach sensitive coverage is represented in an \textit{approach bitmap}, where for every unique $edge\_id$, we set an approach level value, as shown in \fig{sl_eg}. We model this metric in our \MTNN~as an auxiliary task, where the task takes binary test cases as inputs and learn to predict the corresponding approach-level bitmaps.

\subsubsection{Context-sensitive Edge Coverage:~Auxiliary Task 2.~}
\label{sect:ctx}

Edge coverage cannot distinguish between two different test inputs triggering the same edge, but via completely different internal states (\eg, through the same function called from different sites in the program). This distinction is valuable since reaching an edge via a new internal state (\eg, through a new function call site) may trigger a vulnerability hidden deep within the program logic. Augmenting edge coverage with context information regarding internal states of the program may help alleviate this problem~\cite{chen2018angora}.

\begin{figure}[h]
    \centering
    \includegraphics[width=0.9\linewidth]{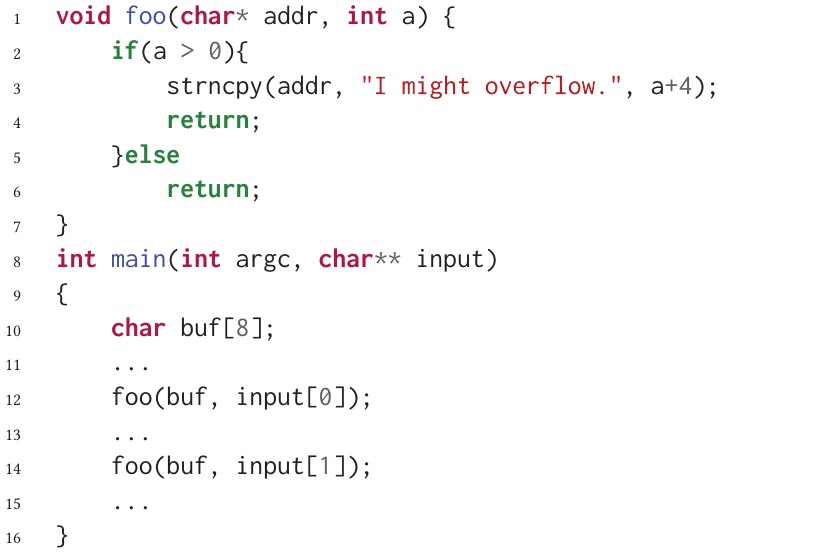}
    \caption{\small An example C-code to demonstrate the usefulness of using context-sensitive measures. Measures such as \ec will fail to detect a possible bug in \texttt{strncpy($\cdot$)}}
    \label{fig:ctx_eg}
\end{figure}

Consider the example 
in \fig{ctx_eg}. Here, for an input \texttt{[1, 0]}, the first call to function \texttt{foo()} appears at site \texttt{line 12} and it triggers the if condition (on \texttt{line 2}); the second call to \texttt{foo()} appears on site \texttt{line 14} and it triggers the else condition (on \texttt{line 5}). As far as \ec is concerned, both the edges of the function \texttt{foo()} (on lines 2 and 5) have been explored and any additional inputs will remain uninteresting. However, if we provide a new input say \texttt{[0, 8]}, we would first trigger \texttt{line 5} of \texttt{foo} when it is called from \texttt{line 12}. Then we trigger \texttt{line 2 } of \texttt{foo} from \texttt{line 14} and further cause a buffer overflow at \texttt{line 3} because a 12 bytes string is written into a 8 bytes destination buffer \texttt{buf}. Moveover, the input \texttt{[0, 8]} will not be saved by \ec fuzzer since it triggers no new edges. Frequently called functions (like \texttt{strcmp}) may be quite susceptible such crashes~\cite{Wang2019}. 

\begin{figure}[h]
    \centering
    \includegraphics[width=0.55\linewidth]{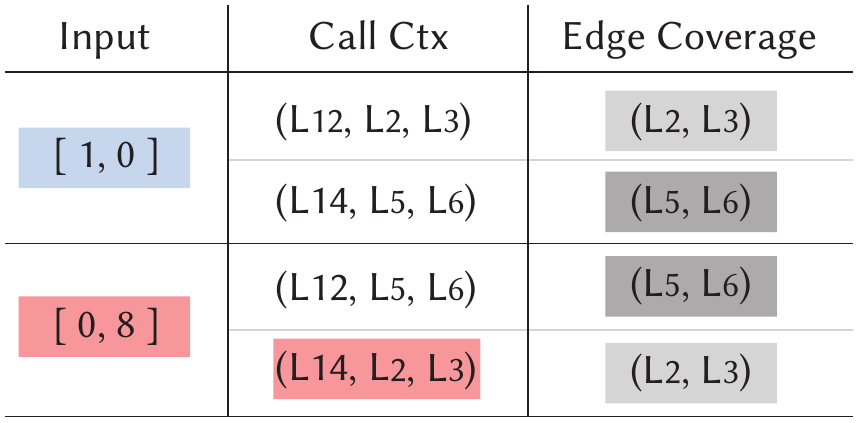}
    \caption{\small The tuple in \ec does not differentiate between the \colorbox{blueish!63}{clean} input and the \colorbox{red!36}{buggy} input, while of \ctx (labeled `Call Ctx') does.}
    \label{fig:ctx_eg_2}
\end{figure}

In order to overcome this challenge, Chen \etal~\cite{chen2018angora} propose keeping track of the \textit{call stack} in addition to the edge coverage by maintaining tuple: 
$(\textit{call\_stack}, \textit{prev\_block}, \textit{cur\_block})$. \fig{ctx_eg_2} shows the additional information provided by \ctx over \ec. Here, we see an example where a buggy input \colorbox{red!36}{\texttt{[0,8]}} has the exact same edge coverage as the clean input \colorbox{blueish!63}{\texttt{[1,0]}}. However, the call context information can differentiate these two inputs based on the call stacks at \texttt{line 12} and \texttt{14}.

We model \ctx in our framework as an auxiliary task. We first assign a unique id to every call. Next, at run time, when we encounter a call at an edge ($edge\_id$), we first compute a $hash$ to record all the functions on current call stack as: $call\_stack = call\_id_1 \oplus ... \oplus call\_id_n$, 
where $call\_id\_i$ represents the \inlinemathit{i}-th function on current call stack and $\oplus$ denotes XOR operation. Next, we compute the \textit{context sensitive edge  id} as: $call\_trace\_id = call\_stack \oplus edge\_id$.
%
%

Thus we obtain a unique $call\_trace\_id$ for every function called from different contexts (\ie, call sites). We then create a bit-map of all the $call\_trace\_id$s. Unlike existing implementations of context-sensitive edge coverage~\cite{chen2018angora, Wang2019}, we assign an additional id to each call instruction while maintaining the original $edge\_id$ intact. Thus, the total number of elements in our bit map reduces to sum of $call\_trace\_id$s and $edge\_id$s 
rather than a product of $call\_trace\_id$s and $edge\_id$s. An advantage of our design is that we minimize the bitmap size. In practice, existing methods requires around $7\times$ larger bitmap size than just \ec~\cite{chen2018angora}; our implementation only requires around $1.3\times$ bitmap size of \ec. The smaller bitmap size can avoid edge explosion and improve performance.

In our multi-tasking framework, the \ctx ``task'' is trained to predict the mapping between the inputs and the corresponding $call\_trace\_id$s  bitmaps. This can enable us to learn the difference between two inputs in a more granular fashion. For example, an ML model can learn that under certain circumstances, the second input byte (\texttt{input[1]} in~\fig{ctx_eg}) can cause crashes. This information cannot be learned by training to predict for \ec alone since both inputs will have the same edge coverage (as shown in~\fig{ctx_eg_2}).

\input{body/3_methodology_stage_1}

\input{body/3_methodology_stage-2_2.tex}

\input{body/3_methodology_stage_3.tex}

%% file: body/3_methodology_stage_1.tex
\subsection{Stage-I: Multi-Task Training}
\label{sect:stage-1}


This phase builds a multi-task neural network (MTNN) that can predict different types of edge coverage given a test input. 
The trained model is designed to produce a more general and compact embedding of the input space focusing only on those  input bytes that are most relevant to all the tasks. This compact representation will be reused by the subsequent stages of the program to identify the most important bytes in the input (\ie, the hot-bytes) and guide mutations on those bytes.

\subsubsection{Architecture.~} \fig{bg_fig_1} shows the architecture of the MTNN. The model contains an \textit{encoder} (shared among all the tasks) and three task-specific \textit{decoders}. 
The model takes existing test input bytes as input and outputs task-specific bitmap. Each input byte corresponds to one input node, and each bitmap value corresponds to an ouput node.

\noindent
\textbf{Encoder.} Comprises of one input layer followed by three progressively narrower intermediate layers. The total number of nodes in the input layer is equal to the total number of bytes in the largest input in the seed corpus. All shorter inputs are padded with \texttt{0x00} for consistency. 
The last layer of the encoder is a compact representation of the input to be used by all the tasks (\colorbox{greenish!80}{green} in \fig{bg_fig_1}). 
\noindent
\textbf{Decoders.} There are three task-specific decoders (shown in \colorbox{lavender}{lilac} in \fig{bg_fig_1}). Each task specific decoder consists of three intermediate layers that grow progressively wider. The last layer of each of the decoder is the output layer. For \ec, there is one node in the output layer for each unique $edge\_id$, likewise for \ctx there is one output node for each $call\_trace\_id$, and for \softlabel there is one output node for each unique $edge\_id$ but they take continuous values (see~\fig{sl_eg}).
\subsubsection{Loss functions.~}
\label{sect:adaptive_loss}

The loss function of a MTNN is a weighted sum of the task-specific loss functions. 
Among our three tasks, \ec and \ctx are modeled as classification tasks and \softlabel is modeled as a regression task. Their loss functions are designed accordingly.


\noindent\textbf{Loss function for \softlabel.} Approach-level measures how close an input was from an edge that was not triggered. This distance is measured using a \textit{continuous value} between 0 and 1. Therefore, this is a regression problem and we use \textit{mean squared error loss}, given by: 
\begin{equation}
\label{eq:mse_loss}
    \mathcal{L}_{\tau_\text{\tiny approach}} = \operatorname{MSE}=\frac{1}{n}\sum_{i=1...n}(Y_i-\hat{Y_i})^2.
\end{equation}
Where $Y_i$ is the prediction and $\hat{Y_i}$ is the ground truth.

\input{tex/background_fig_1.tex}

\noindent\textbf{Loss functions for \ec and \ctx}. The outputs of both these tasks are binary values where \texttt{1} means an input triggered the $edge\_id$ or the $call\_trace\_id$ and \texttt{0} otherwise. We find that while some $edge\_id$s or $call\_trace\_id$s are invoked very rarely, resulting in imbalanced classes. This usually happens when an input triggers a previously unseen (rare) edges. Due to this imbalance, training with an off-the-shelf loss functions such as cross entropy is ill suited as it causes a lot of false negative errors often missing these rare edges. 



To address this issue, we introduce a parameter called \textit{penalty} (denoted by $\beta$) to penalize these false negatives. The penalty is the ratio of the number of times an edge is \textit{not} invoked over the number of times it is invoked. That is, 
{
 \setlength{\abovedisplayskip}{3pt}
 \setlength{\belowdisplayskip}{5pt}
\begin{equation*}
\small
    \text{Penalty} = \beta_\tau = \frac{\text{\# times an edge\_id (or call\_trace\_id) is \textit{not} invoked}}{\text{\# times an edge\_id (or call\_trace\_id) is invoked}}
\end{equation*}
}
Here, $\beta_\tau$ represents the penalty for every applicable task \inlinemath{\tau}$\in$\inlinemathcal{T} and it is dynamically evaluated as fuzzing progresses. Using $\beta_\tau$ we define an \textit{adaptive loss} for classification tasks in our MTNN as:
{
 \setlength{\abovedisplayskip}{5pt}
 \setlength{\belowdisplayskip}{7pt}
\begin{equation}
\label{eq:adap_ent}
    \mathcal{L}_{\tau_\text{\tiny ec/ctx}} = -\sum_{\text{edge}}\left(\beta_\tau\cdot p\cdot log(\hat{p}) + (1 - p)\cdot log(1-\hat{p})\right)
    \vspace{-2mm}
\end{equation}
}
In \eq{adap_ent}, $\mathcal{L}_{\tau_\text{\tiny ec/ctx}}$ results in two separate loss functions for \ec and \ctx. The penalty ($\beta_\tau$) is used to penalize false-positive and false-negative errors. $\beta_\tau>1$ penalizes $p\cdot log({\hat{p}})$, representing false negatives; $\beta_\tau=1$ penalizes both false positives and false negatives equally; and $\beta_\tau<1$ penalizes $(1-p)\cdot log({1 - \hat{p}})$, representing false positives. With this, we compute the total loss for our multi-task NN model with $K$ tasks:
 \begin{equation}
  \setlength{\abovedisplayskip}{3pt}
 \setlength{\belowdisplayskip}{5pt}
     \label{eq:adap_total}
     \displaystyle
     \mathcal{L}_{total} = -\sum_{i=1}^K\alpha_{i}\mathcal{L}_{i} 
 \end{equation}
 This is the weighted sum of the adaptive loss $\mathcal{L}_{i}$ for each individual task. Here, $\alpha_{i}$ presents the weight assigned to task $i$.



%% file: tex/background_fig_1.tex
\begin{figure}[tbp!]
    \centering
    \includegraphics[width=0.95\linewidth]{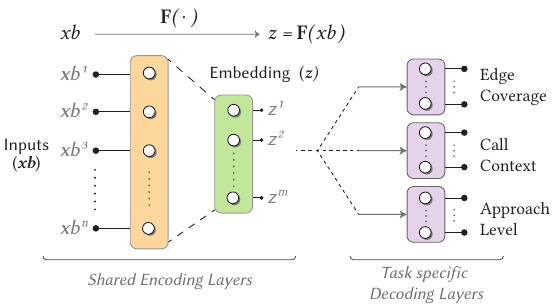}
    \caption{{\small The MTNN architecture representing the n-dimensional \colorbox{orangish!50}{input layer} $\mathbf{xb^i\in xb}$; m-dimensional compact \colorbox{greenish!50}{embedding} layer $\mathbf{z^j\in z}$, s.t. $\mathbf{m<n}$, with a function $\mathbf{F(\cdot)}$ to map input $\mathbf{xb}$ and the embedding layer $\mathbf{xb}$; three \colorbox{lavender!80}{task-specific layers}.}}
    \label{fig:bg_fig_1}
\end{figure}

%% file: body/3_methodology_stage-2_2.tex
\subsection{Stage-II: Guided Mutation}
\label{sect:stage-2}
This phase uses the trained MTNN to generate new inputs that can maximize the code coverage. This is achieved by focusing mutation on the byte locations in the input that can influence the branching behavior of the program (\textit{hot-bytes}). 

We use the the compact embedding layer of the MTNN (shown in \colorbox{greenish!80}{green} in \fig{bg_fig_1}) to infer the hot-byte distribution. The compact embedding layer is well suited for this because it (a)~captures the most semantically meaningful features (\ie, bytes) in the input in a compact 
manner; and (b)~learns to ignore task specific noise patterns~\cite{kumar2017variational} and pays more attention to the important bytes that apply to all tasks~\cite{chen2016, kulkarni2015, Bengio13}.  








Formally, we represent the input (shown in \colorbox{orangish!80}{orange} in \fig{bg_fig_1}) as a byte vector $\mathbf{xb}=\{xb^1, xb^2, ..., xb^n\}\in\mathrm{\left[0, 255\right]^n}$, where $xb^i$ is the $i^{\text{th}}$ byte in $\mathbf{xb}$ and $n$ represents the input dimensions (\ie, number of bytes). 
Then, after the MTNN has been trained, we obtain the compact embedding layer $\mathbf{z}=\{z^1, z^2, ..., z^m\}$ consisting of $\mathit{m}$ nodes. 


Note that when input bytes in $\mathbf{xb}$ changes,  $\mathbf{z}$ changes accordingly. The amount of the change is determined by how influential each byte in the input is to all the tasks in the MTNN model. Changes to the \textit{hot-bytes}, which are more influential, will result in \textit{larger} changes to $\mathbf{z}$. We use this property to discover the \textit{hot-bytes}.

To determine how influential each byte in $\mathbf{xb}$ is, we compute the partial derivatives of the nodes in compact layer with respect to all the input bytes. The partial derivative of the $j$-th node in the embedding layer with respect to the \inlinemathit{i}-th input byte is: 
{
\begin{equation}
\label{eq:partial_d}
\nabla_{\mathbf{xb}}~\mathbf{z}=\left[\dfrac{\partial f_j(\mathbf{x})}{xb^i}\right] = \left[\dfrac{\partial z^j}{\partial xb^i}\right]_{i=1\ldots n,~j=1\dots m}
\end{equation}
}
%
%
In order to infer the importance of each byte $xb^i\in xb$, we define a \textit{saliency score} for each byte, denoted by $\mathbf{\mathrm{S}}(xb^i)$. We compute the saliency score as follows: 
{
\begin{equation}
\label{eq:saliency}
\mathbf{\mathrm{S}}(xb^i) = \mathlarger{\mathlarger{\sum}}_{j=1}^{m}\left|\dfrac{\partial z^j}{\partial xb^i}\right| \hspace{1em} i=1\dots n
\end{equation}
}
The saliency score $\mathbf{\mathrm{S}}(xb^i)$ is the sum of all the partial derivatives in $\nabla_{\mathbf{xb}}~\mathbf{z}$  
\wrt to the byte $xb^i$. The numeric value of each of the $\mathit{n}$ elements in $\mathrm{S}(\mathbf{xb})$ determines the hotness of each byte. The larger the saliency score of an input byte $xb^i$ is, the more likely it is to be a \textit{hot-byte}. 
Using the \textit{saliency score}, 
we can now mutate a test input to generate new ones. 
To do this, we identify the $\mathit{top-k}$ bytes with the largest saliency values\textemdash these are the byte-locations 
that will be mutated by our algorithm. 


For each selected byte-locations, 
we create a new mutated input by changing the bytes to all permissible values between \texttt{0} and \texttt{255}. Since this only happens to the $\mathit{top-k}$ \textit{hot-bytes}, the number of newly mutated seeds remains manageable. We use these mutated inputs for fuzzing and monitor various coverage measures. 




%% file: body/3_methodology_stage_3.tex
\subsection{Stage-III: Seed Selection \& Incremental Learning}
\label{sect:stage-3}

In this step, \tool samples some of the mutated inputs from the previous stage to retrain the model. Sampling inputs is crucial as the choice of inputs can significantly affect the fuzzing performance. Also, as fuzzing progresses, the pool of available inputs keeps growing in size. Without some form of sampling strategy, training the NN and computing gradients would take prohibitively long. 

To this end, we propose an importance sampling~\cite{mcbook} strategy where inputs are sampled such that they reach some {\em important region} of the \cfg instead of randomly sampling from available input. In particular, our sampling strategy first retains all inputs that invoke previously unseen edges. Then, we sort all the seen edges by their rarity. The rarity of an edge is computed by counting how many inputs trigger that specified edge. Finally, we select the top $T$-rarest edges and include at least one input triggering each of these rare edges.
We reason that, by selecting the inputs that invoke the rare edges, we may explore deeper regions of the program on subsequent mutations. In order to limit the number of inputs sampled, we introduce a sampling budget $K$ that determines how many inputs will be selected per iteration.  

Using these sampled inputs, we retrained the model periodically to refine its behavior\textemdash as more data is becoming available about new execution behavior, retraining makes sure the model has knowledge about them and make more informed predictions.

%% file: body/6_experiments.tex
\section{Evaluation}

\input{tex/datasets.tex}

\noindent\textbf{Implementation.~}~
Our MTNN model is implemented in Keras-2.2.3 with Tensorflow-1.8.0 as a backend~\cite{tf, keras}. 
The MTNN is based on a feed-forward model, composed of one shared encoder and three independent decoders. The encoder compresses an input file into a $512$ compact feature vector and feeds it into three following decoders to perform different task predictions. For encoder, we use three hidden layers with dimensions 2048, 1024 and 512. For each decoder, we use one final output layer to perform corresponding task prediction. The dimension of final output layer is determined by different programs. We use ReLU as activation function for all hidden layers. We use sigmoid as the activation function for the output layer. For task specific weights, set each task to equal weight ($\alpha_\tau=1$ in Eq.~\ref{eq:adap_total}). The MTNN model is trained for 100 epochs achieving a test accuracy of around $95\%$ on average. We use Adam optimizer with a learning rate of 0.001. 
As for other hyperparameters, we choose k=1024 for $top-K$ hot-bytes. For seed selection budget $T$ in Stage-III (\tion{stage-3}), we use $T=750$ input samples where each input reaches atleast one rare edge. We note that all these parameters can be tuned in our replication package. 


To obtain the various coverage measures, we implement a custom LLVM pass. Specifically, we instrument the first instruction of each basic block of tested programs to track edge transition between them. We also instrument each call instructions to record the calling context of tested programs at runtime. Additionally, we instrument each branch instructions to measure the distance from branching points to their corresponding descendants. As for magic constraints, we intercept operands of each \texttt{CMP} instruction and use direct-copy to satisfy these constraints.

\noindent\textbf{Study Subjects.~} 
We evaluate \tool on 10 real-world programs, as shown in~\tab{datasets}. To demonstrate the performance of \tool, we compare the edge coverage and number of bugs detected by \tool with $5$ state-of-the-art fuzzers listed in \tab{fuzzers}. Each state-of-the-art fuzzer was run for 24 hours. The training, retraining, and fuzzing times are included in the total 24 runs for each fuzzer. Training time for \tool is shown in~\tab{datasets}. MTFuzz and Neuzz both use the same initial seeds for the the approaches and the same fuzzing backend for consistency. We ensure that all other experimental settings were also identical across all studied fuzzers.

\noindent\textbf{Experimental Setup.~} 
All our measurements are performed on a system running Ubuntu 18.04 with Intel Xeon E5-2623 CPU and an Nvidia GTX 1080 Ti GPU. 
For each program tested, we run AFL-2.52b~\cite{zalewski2017american} on a single core machine for an hour to collect training data. The average number of training inputs collected for $10$ programs is around $2K$. 
We use \texttt{10KB} as the threshold file size for selecting our training data from the AFL input corpus (on average 90\% of the files generated by AFL were under the threshold).

\section{Experimental Results}
\label{sect:rqs}

We evaluate \tool with the following research questions:
\bi[leftmargin=*, wide=0pt]
\item \rqone
\item \rqtwo
\item \rqthree
\item \rqfour
\ei


\subsection*{RQ1: Performance} 
\label{eval:rq1}

We compare \tool wiht other fuzzers (from~\tab{fuzzers}) in terms of the number of real-world and synthetic bugs detected (RQ1-A and RQ1-B), and edge coverage (RQ1-C).

\smallskip\noindent\textit{{\normalsize RQ1-A.~How many real world bugs are discovered by \tool compared to other fuzzers?}}\hfill
\label{sect:rq1}

\noindent\textbf{\textit{Evaluation}.}~To evaluate the number of bugs discovered by a fuzzer, we first instrument the program binaries with \texttt{Address\-Sani\-tize\-r}~\cite{asan} and \texttt{UnderfinedBehaviorSanitizer}~\cite{ubsan}. Such instrumentation is necessary to detect bugs beyond crashes. Next, we run each of the fuzzers for 24 hours (all fuzzers use the same seed corpus) and gather the test inputs generated by each of the fuzzers. We run each of these test inputs on the instrumented binaries and count the number of bugs found in each setting.  
Finally, we use the stack trace of bug reports generated by two sanitizers to categorize the found bugs. Note, if multiple test inputs trigger the same bug, we only consider it once. 
Table~\ref{tab:bug_summary} reports the results.

\begin{figure}[h]
    \centering
    \includegraphics[width=0.9\linewidth]{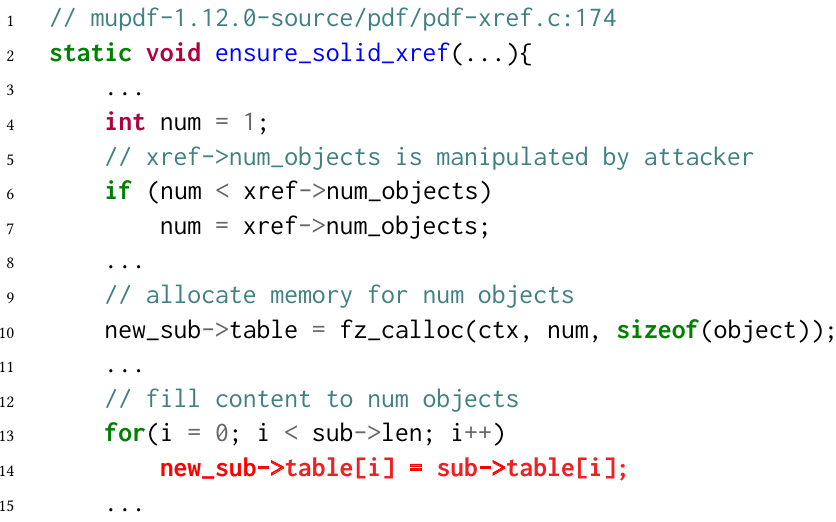}
    \caption{\small Heap overflow bug in \texttt{mupdf}. The \Red{red} line shows the bug.}
    \label{fig:pdf_bug}
\end{figure}

\noindent{{\textbf{Observations.}}}~ We find that:
\begin{enumerate}[leftmargin=*]
\item \tool finds a \textbf{total of 71 bugs}, the most among other five fuzzers in 7 real world programs. In the remaining three programs, no bugs were detected by any fuzzer after 24 hours.
\item Among these, \textbf{11 bugs were previously unreported}. 
\end{enumerate}   
Among the other fuzzers, Neuzz (another ML-based fuzzer) is the second best fuzzer, finding 60 bugs. Angora finds 58.
We observe that the 11 new bugs predominantly belonged to 4 types: memory leak, heap overflow, integer overflow, and out-of-memory. Interestingly, \tool discovered a potentially serious heap overflow vulnerability in \textit{mupdf} that was not found by any other fuzzer so far (see~\Cref{fig:pdf_bug}). 
A mupdf function \texttt{ensure\_solid\_xref} allocates memory (line 10) for each object of a pdf file and fills content to these memory chunks (line 14). Prior to that, at \texttt{line 6}, it tries to obtain the total number of objects by reading a field value \texttt{xref->num\_objects} which is controlled by program input: a pdf file. \tool leverages gradient to identify the hot bytes which control \texttt{xref->num\_objects} and sets it to a \emph{negative} value. As a result, \texttt{num} maintains its initial value $1$ as line 6 if check fails. Thus, at \texttt{line 10}, the function allocates memory space for a single object as $num=1$. However, in \texttt{line 14}, it tries to fill more than one object to \texttt{new\_sub->table} and causes a heap overflow. This bug results in a crash and potential \text{DoS} if \texttt{mupdf} is used in a web server. 

\input{tex/bugs}
\input{tex/RQ5.tex}

\smallskip\noindent\textit{{\normalsize RQ1-B.~How many synthetic bugs in LAVA-M dataset are discovered by \tool compared to other fuzzers?}}\hfill

\noindent\textbf{\textit{Evaluation}.}~LAVA-M is a synthetic bug benchmark where bugs are injected into four GNU \texttt{coreutil} programs~\cite{lava}. Each bug in LAVA-M dataset is guarded by a magic number condition. When the magic verification is passed, the corresponding bug will be triggered. Following conventional practice, we run ~\tool and other fuzzers from~\tab{fuzzers} on LAVA-M dataset for a total of 5 hours. We measure the number of bugs triggered by each of the state-of-the-art fuzzers. The result is tabulated in ~\tab{lava_bug}.


\noindent{{\textbf{Observations.}}}~\tool discovered the most number of bugs on all 4 programs after 5 hours run. The better performance is attributed to the direct-copy module. To find a bug in LAVA-M dataset, fuzzers need to generate an input which satisfies the magic number condition.~\tool's direct-copy module is very effective to solve these magic number verification since it can intercept operands of each \texttt{CMP} instruction at runtime and insert the magic number back into generated inputs.

\input{tex/RQ1.tex}

\input{tex/cov_tab.tex}
\smallskip
\noindent\textit{RQ1-C. How much \ec does \tool achieve compared to other fuzzers?}\hfill

\noindent\textbf{{Evaluation}.}~
To measure edge coverage, we run each of the fuzzers for 24 hours (all fuzzers use the same seed corpus). We periodically collect the edge coverage information of all the test inputs for each fuzzer using AFL’s coverage report toolkit \texttt{afl-showmap}~\cite{zalewski2017american}.
AFL provides coverage instrumentation scheme in two mainstream compilers \texttt{GCC} and \texttt{Clang}. While some authors prefer to use \texttt{afl-gcc}~\cite{she2018neuzz, lemieux2017fairfuzz, bohme2017coverage}, some others use \texttt{afl-clang-fast}\cite{chen2018angora, GREYONE}. The underlying compilers can have different program optimizations which affects how edge coverage is measured. Therefore, in order to offer a fair comparison with previous studies, we measure edge coverage on binaries compiled with with both \texttt{afl-gcc} and \texttt{afl-clang-fast}. In the rest of the paper, we report results on programs compiled with \texttt{afl-clang-fast}. We observed similar findings with \texttt{afl-gcc}.  

\noindent\textbf{{Observations}.}~
The results for edge coverage after 24 hours of fuzzing are tabulated in \tab{cov_main}.
The edge-coverage gained over time is shown in~\fig{RQ1_charts}. 
Overall, \tool achieves noticeably more edge coverage than all other baseline fuzzers. 
Consider the performance gains obtained over the following families of fuzzers:

\noindent\textit{$\circ$~Evolutionary fuzzers}: \tool outperforms all the three evolutionary fuzzers studied here. \tool outperforms Angora on the 6 programs which Angora supports and achieves up to \textbf{$2297$} more edges in \texttt{objdump}. Note, Angora can't run on some programs due to the external library issue on its taint analysis engine~\cite{chen2018angora,Aschermann2019REDQUEENFW}. When compared to both FairFuzz and AFLFast, \tool covers significantly more edges, \eg, $4702$ more than FairFuzz in \texttt{readelf} and over 28.1$\times$ edges compared to AFLFast on~\texttt{objdump}.

\noindent\textit{$\circ$~Machine learning based fuzzers}:~In comparison with the state-of-the-art ML based fuzzer,  Neuzz~\cite{she2018neuzz}, we observed that \tool achieves much greater edge coverage in all 10 programs studied here. We notice improvements of \textbf{2000} more edges in \texttt{readelf} and \textbf{2500} more edges in \texttt{nm} and \texttt{strip}.


\begin{result}
\tool found 71 real-world bugs (11 were previously unknown) and also reach on average $1,277$ and up to $2,867$ more edges compared to Neuzz, the second-best fuzzer, on $10$ programs.
\end{result}
    

\subsection*{RQ2: Contributions of Auxiliary Tasks}
\label{sect:rq2}
    \tool is comprised of an underlying multi-task neural network (MTNN) that contains one primary task (\ec) and two auxiliary tasks namely, \ctx and \softlabel. 
    A natural question that arises is---\textit{How much does each auxiliary task contributes to the overall performance?}

    \noindent{{\textbf{Evaluation}.}}~To answer this question, We study what would happen to the edge coverage when one of the auxiliary tasks is excluded from the \tool. 
    For this, we build four variants of \tool:
    \begin{enumerate}[leftmargin=*, wide=0pt]
        \item \textsl{(EC)}: A single-task NN with only the primary task to predict \ec.
        \item \textsl{(EC, Call Ctx)}: An MTNN with \ec as the primary task and \ctx as the auxiliary task.
        \item \textsl{(EC, Approach)}: An MTNN with \ec as the primary task and \softlabel as the auxiliary task.
        \item \tool: Our proposed model with \ec as the primary task and two auxiliary tasks \ctx and \softlabel.
    \end{enumerate}
     To rule out other confounders, we ensure that each setting shares the same hyper-parameters and the same initial seed corpus. Also, we ensure that all subsequent steps in fuzzing remain the same across each experiment.  
    With these settings, we run each of the above multi-task models on all our programs from \tab{datasets} for 1 hour to record the edge coverage for each of these MTNN models.
    \noindent{{\textbf{Observations}.}}~Our results are tabulated in~\tab{multi_vs_single}.
    We make the following noteworthy observations:
    \input{tex/multi_vs_single}

    \noindent 1. Fuzzer that uses an MTNN trained on \ec as the primary task and \ctx as the only auxiliary task tends to perform only marginally better than a single task NN based on \ec. In some cases, \eg, in~\tab{multi_vs_single} we notice about 25\% more edges. However, in some other cases, for example \texttt{libjpeg}, we noticed that the coverage reduces by almost $31\%$.

    \noindent 2. The above trend is also observable for using \ec with \softlabel as the auxiliary. For example, in \texttt{libjpeg}, the edge coverage is lower than the single-task model that uses only \ec.
    
    \noindent 3. However, \tool, which uses both \ctx and \softlabel as auxiliary tasks to \ec, performs noticeably better than all other models with up to \textbf{800 more edges covered} ($\mathbf{\approx}$\textbf{20\%}) in the case of \texttt{readelf}. 
    
   The aforementioned behavior is expected because each auxiliary task provides very specific albeit somewhat partial context to \ec. Context-sensitive edge coverage only provides context to triggered edges, while \softlabel only reasons about non-triggered edges (see \tion{tasks} for details). Used in isolation, a  partial context does not have much to offer. However, while working together as auxiliary tasks along with the primary task, it provides a better context to \ec resulting in overall increased edge coverage (see the last column of~\tab{multi_vs_single}).
        
    \begin{result}
        \tool benefits from \textit{both} the auxiliary tasks. Using \ctx and \ec along with the primary task (of predicting edge coverage) is most beneficial. We achieve up to \textbf{20\%} more edge coverage.  
    \end{result}

\subsection*{RQ3. Impact of Design Choices}


    
    
    While building \tool, we made few key design choices such as using a task-specific adaptive loss (\tion{adaptive_loss}) to improve the quality of the multi-task neural network (MTNN) model and a novel seed selection strategy based on importance sampling (see~\tion{stage-3}). Here we assess how helpful these design choices are. 
    
    \smallskip
    \noindent\textit{RQ3-A. What are the benefits of using adaptive loss?}
    \smallskip
    

    \noindent MTNN model predicting for edge coverage and for \ctx tends to experience severely imbalanced class labels. Consider the instance when a certain input triggers an edge for the first time. This is an input of much interest because it represents a new behaviour. The MTNN model must learn what lead to this behaviour. However, in the training sample, there exists only one positive sample for this new edge in the entire corpus. An MTNN that is trained with an off-the-shelf loss functions is  likely to misclassify these edges resulting in a false negative error.
    Such false negatives are particularly damaging because a large number of new edge discoveries go undetected affecting the overall model performance. To counter this, we defined an adaptive loss in \tion{adaptive_loss}; here we measure how much it improves the MTNN's performance.
    
    \noindent{{\textbf{Evaluation}.}}~
    To evaluate the effect of class imbalance, we measure \textit{recall} which is high when the overall false negatives (FN) are low. While attempting to minimize FNs the model must not make too many false positive (FP) errors. Although false positives are not as damaging as false negatives, we must attempt to keep them low. We therefore also keep track of the F1-scores which quantify the trade-off between false positives and false negatives. We train \tool with two different losses (\ie, with our adaptive loss and with the default cross-entropy loss) on $10$ programs for $100$ epochs and record the final recall and F-1 scores.

    \noindent{{\textbf{Observations}.}} The result are shown in~\tab{rq3}. We observe that adaptive loss results in MTNNs with an average of $\mathbf{90\%}$ recall score on $10$ programs, while the default loss model only achieves on average $75\%$ recall score. Generally, we notice \textbf{improvements greater than} $\mathbf{15\%}$ over default loss functions. The low recall for default loss function indicates that it is susceptible to making a lot of false negative predictions. However, our adaptive loss function is much better at reducing false negative predictions. Also, 
    the adaptive loss model achieves on average F-1 score of $72\%$, while unweighted loss model achieves an average of $70\%$. This is encouraging because even after significantly reducing the number of false negatives, we maintain the overall performance of the MTNN.
    
    \begin{result}
        Weighted loss improves \tool's recall by more than 15\%.
    \end{result}

    \smallskip\noindent\textit{RQ3-B. How does seed-selection help?}\smallskip
    
    \noindent{{\textbf{Evaluation}.}}
    Here, we evaluate our seed selection strategy (\tion{stage-3}) by comparing it to a random selection strategy. Specifically, we run two variants of \tool, one with importance sampling for seed selection and the other with a random seed selection. All other components of the tool such as MTNN model, hyperparameters, random seed, etc. are kept constant. We measure the \ec obtained by both the strategies on 10 programs after fuzzing for one hour. \tab{rq3} shows the results.

    \noindent{{\textbf{Observations}.}} When compared to a random seed selection strategy. Importance sampling outperforms random seed selection in all $10$ programs offering average improvements of $1.66\times$ more edges covered than random seed selection\textemdash   
    for \texttt{readelf}, it covers around $2000$ more edges. 
    This makes intuitive sense because, the goal of importance sampling was to retain the newly generated inputs that invoke certain rare edges. By populating the corpus with such rare and novel inputs, the number of newly explored edges would increase over time, resulting in increase edge coverage (see \tab{rq3}).  

    \input{tex/F1}

\input{tex/RQ3.tex}

    \begin{result}
        Importance sampling helps \tool achieve on average $1.66\times$ edge coverage compared with random seed selection.
    \end{result}

    \subsection*{RQ4. Transferability}
    \noindent In this section, we explore the extent to which \tool can be generalized across different programs operating on the same inputs (\eg, two ELF fuzzers). Among such programs, we study if we can transfer inputs generated by fuzzing from one program to trigger edge coverage in another program 
    (RQ4-A) and if it is possible to transfer the shared embedding layers between programs 
    (RQ4-B).

    \smallskip\noindent\textit{RQ4-A. Can inputs generated for one program be transferred to other programs operating on the same domain?}\smallskip
    
    
    \noindent\tool mutates the hot-bytes in the inputs to generate additional test inputs. These hot-bytes are specific to the underlying structure of the inputs. Therefore, inputs that have been mutated on these hot-bytes should be able to elicit new edge coverage for any program that parses the same input. 

\noindent
    {\textbf{Evaluation.}}~To answer this question, we explore 5 different programs that operate on 2 file types: (1) \texttt{readelf}, \texttt{size}, and \texttt{nm} operating on ELF files, and (2)~\texttt{libxml} and \texttt{xmlwf}~\cite{xmlwf} operating on XML files. For all the programs that operate on the same file format:
    
\begin{enumerate}[leftmargin=*]
    \item We pick a source program (say $\mathrm{S}=P_i$) and use \tool to fuzz the source program for 1 hour to generate new test inputs.
    \item Next, for every other target program $\mathrm{T}=P_{j\neq i}$, we use the test inputs from the previous step to measure the coverage. Note that we \textit{do not} deploy the fuzzer on the target program, we merely measure the code coverage.
    \item For comparison, we use \neuzz (another ML-based fuzzer) and AFL to fuzz the source program $\mathrm{S}$ to generate the test inputs for the target program. 
\end{enumerate}
    
    \input{tex/RQ4}

    \noindent
    {\textbf{Observation.}}~We observe from~\tab{rq4} that inputs generated by \tool produce much higher edge coverage on the target program compared to seeds generated by Neuzz or AFL. In general, we notice on average $10\times$ more edge coverage than AFL and $2\times$ more edge coverage than Neuzz.  
    Here, AFL performs the worse, since it generates seeds very specific to the source program. \neuzz, a machine learning based fuzzer, performs better than AFL since it attempts to learn some representation of the input, but it falls short of \tool which learns the most general input representation.  
    
    \smallskip\noindent\textit{RQ4-B. Can the shared layer be transferred between programs?}\smallskip
    
    \noindent We hypothesize that since \tool can learn a general compact representation of the input, it should, in theory, allow for these compact representations to be \textit{transferred} across programs that share the same input, \eg, across programs that process \texttt{ELF} binaries. 
    
    \noindent\textit{\textbf{Evaluation:}}~To verify this, we do the following:
    
    \begin{enumerate}[leftmargin=*]
    \item We pick a source program (say $\mathrm{S}=P_i$) and use \tool to fuzz the source program for 1 hour to generate new tests inputs.
    \item For every target program $\mathrm{T}=P_{j\neq i}$, we transfer the shared embedding layer along with the test inputs from the source program to fuzz the target program. 
    \item Note that the key distinction here is, unlike RQ4-A, here we \textit{fuzz} the target program with \tool using the shared layers and the seed from the source program to bootstrap fuzzing.   
    \end{enumerate}
    
    \noindent{\textbf{Observation.}}~We achieve significantly more edge coverage by transferring both the seeds and the shared embedding layers from the source to target program (\tab{rq4}). On average, we obtain $2\times$ more edge coverage on all $10$ programs. Specifically, transferring the shared embedding layers and the seeds from \texttt{nm} to \texttt{readelf} results in covering \textbf{2$\times$} \textbf{more} edges compared to Neuzz and over \textbf{15}$\times$ \textbf{more} edges compared to AFL. Transferring offers better edge coverage compared to fuzzing the target program with AFL.
    
    \begin{result}
        \tool's compact embedding can be transferred across programs that operate on similar input formats. We achieve up to $14$ times edge coverage for XML files (with an average of $2$ times edge coverage across all programs) compared to other fuzzers. 
    \end{result}

%% file: tex/datasets.tex
\begin{table}
        \begin{minipage}{\linewidth}
            \footnotesize
            \caption{\small Test programs used in our study}
            \label{tab:datasets}
            \resizebox{\linewidth}{!}{
            \begin{tabular}[t]{llrrr}
                \toprule
                \multicolumn{2}{c}{\bf Programs} & \multirow{2}{*}{\bf \# Lines~} & \multicolumn{1}{c}{\multirow{2}{*}{\begin{tabular}[c]{@{}c@{}}\textbf{\tool}\\ \textbf{train (s)}\end{tabular}}} & \multicolumn{1}{c}{\bf Initial  coverage} \\ \cmidrule{1-2} 
                Class & Name &  & \multicolumn{1}{r}{} &  \\ \toprule
                \multirow{5}{*}{\begin{tabular}[l]{@{}l@{}}binutils-2.30~~\\ ELF\\ Parser\end{tabular}} 
                 & readelf -a & 21,647  & 703 & 3,132  \\
                 & nm -C & 53,457 & 202 & 3,031 \\
                 & objdump -D& 72,955 & 703 & 3,939 \\
                 & size & 52,991 & 203 & 1,868  \\
                 & strip & 56,330 & 402 & 3,991  \\ 
                TTF & harfbuzz-1.7.6 & 9,853 & 803 & 5,786  \\ 
                JPEG & libjpeg-9c & 8,857 & 1403 & 1,609  \\ 
                PDF & mupdf-1.12.0 & 123,562 & 403 & 4,641  \\ 
                XML & libxml2-2.9.7 & 73,920 & 903 & 6,372  \\ 
                ZIP & zlib-1.2.11 & 1,893 & 107 & 1,438 \\ \toprule
            \end{tabular}}
            \end{minipage}
            
                \begin{minipage}{\linewidth}
        \caption{\small State-of-the art fuzzers used in our.}
        \label{tab:fuzzers}
        \footnotesize
        \resizebox{\linewidth}{!}{
            \begin{tabular}[t]{rp{0.7\columnwidth}}
                \toprule
                \textbf{Fuzzer} & \textbf{Technical Description} \\
                \toprule
                \afl~\cite{zalewski2017american} & evolutionary search \\
                \aflfast~\cite{bohme2017coverage} & evolutionary + markov-model-based search\\
                FairFuzz~\cite{lemieux2017fairfuzz} & evolutionary + byte masking \\
                Angora~\cite{chen2018angora} & evolutionary + dynamic-taint-guided + coordinate descent + type inference \\
                Neuzz~\cite{she2018neuzz}  & Neural smoothing guided fuzzing \\     
                \toprule
                \end{tabular}
                }
        \end{minipage}
        
\end{table}

%% file: tex/bugs.tex
\begin{table}[t!]
    \caption{\small Real-world bugs found after 24 hours by various fuzzers. \tool finds the most number of bugs, \ie, 71 (11 unseen) comprised of 4 heap-overflows, 3 Memory leaks, 2 integer overflows, and 2 out-of-memory bugs.}
    \label{tab:bug_summary}
    \small
    \arrayrulecolor{gray95}
    \resizebox{\linewidth}{!}{%
    \begin{tabular}{|l|rrrrr|r|}
        \hlineB{2}
        \rowcolor[HTML]{E1E1E1} 
        \textbf{Program} &
          \multicolumn{1}{l}{\cellcolor[HTML]{E1E1E1}\textbf{AFLFast}} &
          \multicolumn{1}{l}{\cellcolor[HTML]{E1E1E1}\textbf{AFL}} &
          \multicolumn{1}{l}{\cellcolor[HTML]{E1E1E1}\textbf{FairFuzz}} &
          \multicolumn{1}{l}{\cellcolor[HTML]{E1E1E1}\textbf{Angora}} &
          \multicolumn{1}{l}{\cellcolor[HTML]{E1E1E1}\textbf{Neuzz}} &
          \multicolumn{1}{|l|}{\cellcolor[HTML]{E1E1E1}\textbf{\tool}} \\\hlineB{2}
        readelf &
          5 &
          4 &
          5 &
          16 &
          16 &
          \cellcolor{blueish!15}17 \\
        {nm} &
          {7} &
          {8} &
          {8} &
          {10} &
          {9} &
          {\cellcolor{blueish!15}12} \\
        objdump &
          6 &
          6 &
          8 &
          5 &
          8 &
          \cellcolor{blueish!15}9 \\
        {size} &
          {4} &
          {4} &
          {5} &
          {7} &
          {6} &
          {\cellcolor{blueish!15}10} \\
        strip &
          5 &
          7 &
          9 &
          20 &
          20 &
          \cellcolor{blueish!15}21 \\
        libjpeg &
          0 &
          0 &
          0 &
          0 &
          1 &
          \cellcolor{blueish!15}1 \\
        mupdf &
          0 &
          0 &
          0 &
          0 &
          0 &
          \cellcolor{blueish!15}1 \\\hlineB{2}
        \textbf{Total} &
          27 &
          29 &
          35 &
          58 &
          60 &
          \cellcolor{blueish!15}71 \\\hlineB{2} 
        \end{tabular}%
        }
        \end{table}

%% file: tex/RQ5.tex
\begin{table}[t!]
    \caption{\small Synthetic bugs in LAVA-M dataset found after 5 hours.
    }
    \label{tab:lava_bug}
    \small
    \arrayrulecolor{gray95}
    \resizebox{0.72\linewidth}{!}{%
    \begin{tabular}{|l|rrr|r|}
        \hlineB{2}
        \rowcolor[HTML]{E1E1E1} 
        \textbf{Program} &
          \multicolumn{1}{l}{\cellcolor[HTML]{E1E1E1}\textbf{\#Bugs}} &
          \multicolumn{1}{l}{\cellcolor[HTML]{E1E1E1}\textbf{Angora}} &
          \multicolumn{1}{l}{\cellcolor[HTML]{E1E1E1}\textbf{Neuzz}} &
          \multicolumn{1}{|l|}{\cellcolor[HTML]{E1E1E1}\textbf{\tool}} \\\hlineB{2}
        base64 &
           44 &
          48 &
          48 &
          48 
          \cellcolor{blueish!15} \\
        md5sum &
          57 &
          57 &
          60 &
          60 
          {\cellcolor{blueish!15}} \\
        uniq &
          28 &
          29 &
          29 &
          29 
          {\cellcolor{blueish!15}} \\
        who &
         2136 & 
          1541 &
          1582 &
          1833 
          \cellcolor{blueish!15} \\
          \hlineB{2}
        \end{tabular}%
        }
        \end{table}

%% file: tex/RQ1.tex
\begin{figure*}[tbp!]
    \includegraphics[width=0.9\linewidth]{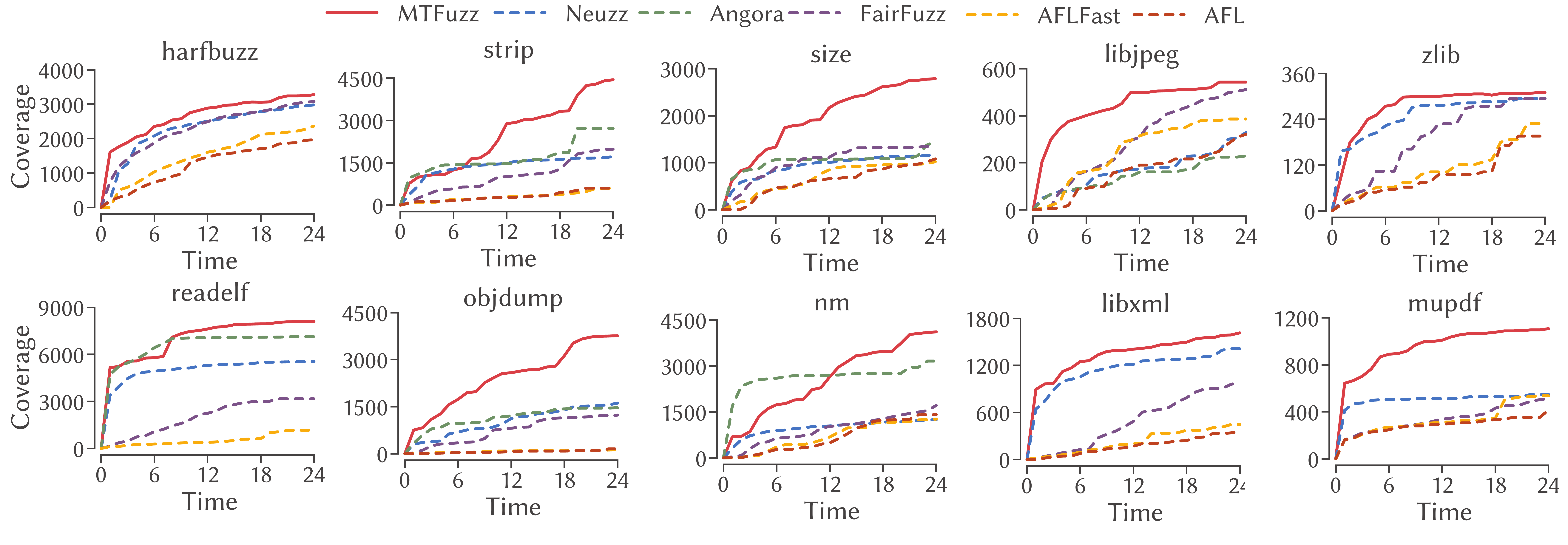}
    \caption{{\small Edge coverage over 24 hours of fuzzing by \tool and other state-of-the-art fuzzers. 
    }}
    \label{fig:RQ1_charts}
\end{figure*}

%% file: tex/cov_tab.tex
\begin{table}[t!]
    \caption{{\small The \textit{average} edge coverage of \tool compared with other fuzzers after 24 hours runs for 5 repetitions. Parenthesized numbers represent the standard deviation.}}
    \label{tab:cov_main}
\begin{subfigure}[t]{\linewidth}
    \caption{Program binaries compiled with \texttt{afl-clang-fast}}
    \label{tab:afl_cov_main}
    \resizebox{\linewidth}{!}{%
    \begin{tabular}{|l|r|rrrrr|}
        \hlineB{2}\rowcolor[HTML]{E1E1E1} 
        \textbf{Program} & \textbf{\tool}  & \textbf{Neuzz} & \textbf{Angora} & \textbf{FairFuzz}& \textbf{\afl} & \textbf{\aflfast}\\ 
            \hlineB{1.5}
            
            readelf  & \cellcolor{blueish!15} 8,109 & 5,953 & 7,757 &3,407 &  701 & 1,232 \\
            
             & \cellcolor{blueish!15} \small\textit{(286)} & \small\textit{(141)} & \small\textit{(248)} & \small\textit{(1005)} &  \small\textit{(87)} & \small\textit{(437)} \\

            nm  & \cellcolor{blueish!15} 4,112 & 1,245 & 3,145 & 1,694 & 1,416 & 1,277 \\
            
             & \cellcolor{blueish!15} \small\textit{(161)} & \small\textit{(32)} & \small\textit{(2033)} & \small\textit{(518)} &  \small\textit{(144)} & \small\textit{(18)} \\

            objdump  & \cellcolor{blueish!15}3,762 &  1,642& 1,465 & 1,225 & 163 &  134 \\
            
             & \cellcolor{blueish!15} \small\textit{(359)} & \small\textit{(77)} & \small\textit{(148)} & \small\textit{(316)} &  \small\textit{(25)} & \small\textit{(15)} \\

            size & \cellcolor{blueish!15}2,786 &1,170  &  1,586& 1,350 & 1,082 & 1,023 \\
            
             & \cellcolor{blueish!15} \small\textit{(69)} & \small\textit{(45)} & \small\textit{(204)} & \small\textit{(47)} &  \small\textit{(86)} & \small\textit{(117)} \\

            strip & \cellcolor{blueish!15}4,406 & 1,653 & 2,682 & 1,920 & 596 & 609 \\
            
             & \cellcolor{blueish!15} \small\textit{(234)} & \small\textit{(110)} & \small\textit{(682)} & \small\textit{(591)} &  \small\textit{(201)} & \small\textit{(183)} \\

            libjpeg & \cellcolor{blueish!15}543 & 328 & 201 & 504 & 327 & 393 \\
            
             & \cellcolor{blueish!15} \small\textit{(19)} & \small\textit{(34)} & \small\textit{(42)} & \small\textit{(101)} &  \small\textit{(99)} & \small\textit{(21)} \\

            libxml & \cellcolor{blueish!15}1,615 &1,419  & $\cdot$ & 956 & 358 & 442 \\
            
             & \cellcolor{blueish!15} \small\textit{(28)} & \small\textit{(76)} & $\cdot$ & \small\textit{(313)} &  \small\textit{(71)} & \small\textit{(41)} \\

            mupdf & \cellcolor{blueish!15}1,107 & 533 & $\cdot$ & 503 & 419 & 536 \\
            
             & \cellcolor{blueish!15} \small\textit{(26)} & \small\textit{(76)} & $\cdot$ & \small\textit{(76)} &  \small\textit{(28)} & \small\textit{(30)} \\

            zlib & \cellcolor{blueish!15}298 & 297 & $\cdot$ & 294 & 196 & 229 \\
            
             & \cellcolor{blueish!15} \small\textit{(38)} & \small\textit{(44)} & $\cdot$ & \small\textit{(95)} &  \small\textit{(41)} & \small\textit{(53)} \\

            harfbuzz & \cellcolor{blueish!15}3,276 & 3,000 & $\cdot$ & 3,060 & 1,992 & 2,365 \\
             
             & \cellcolor{blueish!15} \small\textit{(140)} & \small\textit{(503)} & $\cdot$ & \small\textit{(233)} &  \small\textit{(121)} & \small\textit{(147)} \\
        \hlineB{2}
        \end{tabular}}
        {\scriptsize $\mathbf{\cdot}$ indicates cases where Angora failed to run due to the external library issue.}

\end{subfigure}\\[1em]

\begin{subfigure}[t]{\linewidth}
\caption{Program binaries compiled with \texttt{alf-gcc}}
    \label{tab:afl_cov_gcc}
    \resizebox{\linewidth}{!}{%
    \begin{tabular}{|l|r|rrrrr|}
\hlineB{2}\rowcolor[HTML]{E1E1E1} 
\textbf{Programs} & \textbf{\tool}  & \textbf{Neuzz} & \textbf{Angora} & \textbf{FairFuzz}& \textbf{\afl} & \textbf{\aflfast}\\ 
\hlineB{1.5}
readelf  & \cellcolor{blueish!15} 6,701  & 4,769 & 6,514 &3,423 &  1,072 & 1,314 \\
nm  & \cellcolor{blueish!15} 4,457 & 1,456 & 2,892 & 1,603 & 1,496 & 1,270 \\
objdump  & \cellcolor{blueish!15}5,024 & 2,017& 1,783 & 1,526 & 247 & 187\\
size & \cellcolor{blueish!15}3,728 & 1,737  & 2,107 & 1,954 & 1,426 & 1,446 \\
strip & \cellcolor{blueish!15}6,013 &2,726 & 3,112 & 3,055 & 764 & 757 \\
libjpeg & \cellcolor{blueish!15}1,189 & 719 & 499 & 977 & 671  & 850 \\
libxml & \cellcolor{blueish!15}1,576 &1,357 & $\cdot$ & 1,021& 395 & 388 \\
mupdf & \cellcolor{blueish!15}1,107 & 533 & $\cdot$ & 503 & 419 & 536 \\
zlib & \cellcolor{blueish!15}298 & 297 & $\cdot$ & 294 & 196 & 229 \\
harfbuzz & \cellcolor{blueish!15}6,325 & 5,629 & $\cdot$ & 5,613 & 2,616  & 3,692 \\
\hlineB{2}
\end{tabular}
}

\end{subfigure}
\end{table}

%% file: tex/multi_vs_single.tex
\begin{table}[tb!]
\small
\caption{\small Edge coverage after 1 hour. The most improvement to edge coverage is observed when including both the auxiliary tasks are trained together as a multi-task learner.}
\label{tab:multi_vs_single}
\resizebox{\linewidth}{!}{
\begin{tabular}[t]{|l|rrr|r|}
\hlineB{2}
\bf {\cellcolor{gray05} Programs~} & \cellcolor{gray05} \bf {EC} & \cellcolor{gray05} \bf {EC, Call Stack} & \cellcolor{gray05} \bf {EC, Approach} & \cellcolor{gray05} \bf \tool\\
     \hlineB{1.5}
     readelf  & 4,012 & 4,172 & 4,044 & \cellcolor{blueish!15}4,799\\ 
     nm       & 546  & 532  & 412 &  \cellcolor{blueish!15}577  \\ 
     objdump  &  605 & 632  & 624 & {\cellcolor{blueish!15}{672}} \\ 
     size     & 350  & 404  & 500 & {\cellcolor{blueish!15}{502}} \\ 
     strip    & 744  & 787  & 902 & {\cellcolor{blueish!15}{954}} \\  
     harfbuzz & 593  & 661  & 752 & {\cellcolor{blueish!15}{884}}  \\  
     libjpeg  & 190  & 135  & 182 & {\cellcolor{blueish!15}{223}}  \\ 
     mupdf    & 252  & 193  & 257 & {\cellcolor{blueish!15}{269}} \\ 
     libxml2  & 525  & 649  & 677 & {\cellcolor{blueish!15}{699}} \\ 
     zlib     & 56   & 33   & 59 & {\cellcolor{blueish!15}{67}} \\ 
     \hlineB{2}
\end{tabular}}

\end{table}

%% file: tex/F1.tex

%% file: tex/RQ3.tex
\begin{table}[t]
\caption{\small Impact of design choices. Adaptive loss~(\tion{adaptive_loss}) increases Recall by $\mathbf{\sim 15\%}$ while maintaining similar F1-scores. Seed selection based in importance sampling~(\tion{stage-3}) demonstrate notable gains in overall edge coverage.}
\label{tab:rq3}
\small
\center
\resizebox{\linewidth}{!}{
\begin{tabular}{|l|>{\columncolor{blueish!25}} r|>{\columncolor{blueish!25}}r| r|r||>{\columncolor{blueish!25}} r|r|}
\hlineB{2}
\cellcolor[HTML]{EFEFEF} & \multicolumn{2}{c|}{\cellcolor[HTML]{EFEFEF}Adaptive} & \multicolumn{2}{c||}{\cellcolor[HTML]{EFEFEF}Default} & \multicolumn{2}{c|}{\cellcolor[HTML]{EFEFEF}Seed Selection} \\
\cellcolor[HTML]{EFEFEF}\multirow{-2}{*}{Programs} & \cellcolor[HTML]{EFEFEF}Recall(\%) & \cellcolor[HTML]{EFEFEF}F1(\%) & \cellcolor[HTML]{EFEFEF}Recall(\%) & \cellcolor[HTML]{EFEFEF}F1(\%) & \cellcolor[HTML]{EFEFEF}Our Approach & \cellcolor[HTML]{EFEFEF}Random \\
\hlineB{2}
readelf & 88 & 68 & 74 & 66 & 4,799 & 2,893  \\
nm & 89 & 62 & 69 & 62 & 577 & 269 \\
objdump & 89 & 72 & 65 & 71 & 672 & 437 \\
size & 94 & 81 & 78 & 78 & 502 & 312 \\
strip & 89 & 73 & 80 & 72 & 954 & 545 \\
harfbuzz & 92 & \cellcolor{blueish!25}67 & 80 & 71 & 884 & 558 \\
libjpeg & 88 & 68 & 65 & 65 & 223 & 124 \\
mupdf & 92 & 84 & 90 & 84 &  269 & 160 \\
libxml2 & 90 & 70 & 76 & 69 & 699 & 431 \\
zlib   & 86 & 70 &70 & 65 & 67 & 57\\
\hlineB{2}
\end{tabular}}
\end{table}

%% file: tex/RQ4.tex





\begin{table}[t!]
    \caption{\small Generalizability of \tool across different programs parsing the same file types (\texttt{ELF} and \texttt{XML}). The numbers shown represent new edge coverage.}
    \label{tab:rq4}
    \centering
    \small
    \arrayrulecolor{black}
\resizebox{\linewidth}{!}{%
\begin{tabular}{|l|l|
>{\columncolor{blueish!25}}r |rrr|}
\hline
\cellcolor{gray05}          & \cellcolor[HTML]{EFEFEF}                                              & \cellcolor[HTML]{EFEFEF}Inputs + Embedding  & \multicolumn{3}{c|}{\cellcolor[HTML]{EFEFEF}Inputs only (RQ4-A)}                                    \\ \cline{3-6} 
\multirow{-2}{*}{\cellcolor[HTML]{EFEFEF}File Type} & \multirow{-2}{*}{\cellcolor[HTML]{EFEFEF}Source $\rightarrow$ Target} & \multicolumn{1}{c|}{\cellcolor[HTML]{EFEFEF}\tool  (RQ4-B)}            & \cellcolor[HTML]{EFEFEF}\tool & \cellcolor[HTML]{EFEFEF}Neuzz & \cellcolor[HTML]{EFEFEF}AFL \\ \hline
 & nm$\rightarrow$ nm* & 668 & 668  & 315 & 67    \\
 & size $\rightarrow$ nm  & 312 & 294   & 193  & 32 \\
 & readelf $\rightarrow$ nm & 185& 112   & 68   & 13   \\

 & size $\rightarrow$ size*  & 598  & 598 & 372 & 46 \\
 & readelf $\rightarrow$ size & 218   & 151  & 87    & 19    \\
 & nm$\rightarrow$ size & 328 & 236   & 186   & 17 \\

 & readelf $\rightarrow$ readelf*  & 5,153  & 5,153 & 3,650 & 339 \\
 & size $\rightarrow$ readelf  & 3,146 & 1,848   & 1,687   & 327  \\
\multirow{-9}{*}{ELF}  & nm $\rightarrow$ readelf  & 3,329 & 2,575 & 1,597  & 262 \\\hline 
& xmlwf $\rightarrow$ xmlwf* & 629  & 629 & 343  & 45 \\
& libxml2 $\rightarrow$ xmlwf & 312  & 304  & 187  & 19 \\

& libxml2 $\rightarrow$ libxml2* & 891  & 891  & 643  & 73 \\
\multirow{-4}{*}{XML}  & xmlwf $\rightarrow$ libxml2 & 381   & 298   & 72  & 65  \\ \hline
\end{tabular}
}
{\scriptsize * indicates baseline setting without transfer learning}

\end{table}

%% file: body/7_discussions.tex
\section{Threats to validity}
\label{sect:threats}
(a)~%
\noindent{\em Initialization}: For the fuzzers studied here, it is required to provide initial set of seed inputs. To ensure a fair comparison, we use the same set of seed inputs for all the fuzzers. 
\\(b)~%
\noindent{\em Target programs}: 
We selected diverse target programs from a wide variety of software systems. One still has to be careful when generalizing to other programs not studied here. We ensure that all the target programs used in this study have been used previously; we do not claim that our results generalize beyond these programs. 
\\(c)~%
\noindent{\em Other fuzzers}: When comparing \tool with other state-of-the-art fuzzers, we use those fuzzers that are reported to work on the programs tested here. Our baseline fuzzer Neuzz~\cite{she2018neuzz} has reported to outperform many other fuzzers on the same studied programs. Since we are outperforming Neuzz, it is reasonable to expect that we will outperform the other fuzzers as well. 



%% file: body/8_related.tex
\section{Related Work}
\label{sect:related}
Fuzzing~\cite{miller1990empirical} has garnered significant attention recently. There are three broad types of fuzzers: (a)~ Blackbox ~\cite{hocevar2011zzuf, hoschele2016dynamic, cha2015program} with no knowledge of the target program, (b)~Whitebox~\cite{cadar2008klee, godefroid2005dart, sen2006cute, godefroid2008automated} with source/binary level access the target program, and (c) Greybox fuzzers like AFL with the ability to instrument and collect some target-program-specific information like code coverage. 
This paper specifically focuses on greybox fuzzers. Most greybox fuzzers use evolutionary search to guide their input generation strategy~\cite{zalewski2017american}. Since the release of AFL~\cite{zalewski2017american}, the researchers have attempted to implement a wide range of mutation strategies augmented with program analysis to optimize the evolutionary mutation process~\cite{bohme2017coverage, lemieux2017fairfuzz, You2019ProFuzzerOI,You2019SLFFW,chen2018angora,Aschermann2019REDQUEENFW, grimoire2019, lemieux2017fairfuzz, aflsmart}. All of these projects focus on manually designing different mutation strategies and use either program analysis~\cite{chen2018angora,Aschermann2019REDQUEENFW, grimoire2019} or aggregate statistics~\cite{lemieux2017fairfuzz} to customize their strategy for specific target programs. By contrast, \tool uses multi-task neural networks to automatically learn an compact representation of input to identify and mutate the hot-bytes.  

More recently, machine learning techniques are being increasingly used to improve fuzzing. One line of work focused on using neural networks to model the input format of the target programs based on a corpus of sample inputs~\cite{godefroid2017learn, Rajpal2017, bastani2017synthesizing, bottinger2018deep, she2018neuzz}. Another alternative approach like Neuzz~\cite{she2018neuzz} models the edge behaviour of a program using a neural network. In this paper, we demonstrate that neural networks can be further used to adaptively learn a number of mutation parameters that can significantly improve edge coverage.


Transfer learning~\cite{Pratt1992Dis, Hell2009Feature, Silver2008Guest,Finn2015DeepSA,Mihal2007Map} is beneficial when there is insufficient data for a target task but there exists sufficient data for other source task. To ensure ideal transfer, the target task and source task should have same or similar feature space and share similar data distribution. \citeauthor{Raina2006Con}~\cite{Raina2006Con} and \citeauthor{Dai2007Co}~\cite{Dai2007Co}~\cite{Dai2007Trans} use  transfer learning to perform cross-domain text classification. \citeauthor{Long2015Learn}~\cite{Long2015Learn} and \citeauthor{Sun2016DeepCC}~\cite{Sun2016DeepCC} apply transfer learning to solve image-classification problem. We demonstrate that \tool can transfer a NN learnt on one program to other similar programs.

%% file: body/9_conclusion.tex
\section{Conclusion}
\label{sect:conclusion}

This paper presents \tool, a multi-task neural-network fuzzing framework. \tool learns from multiple code coverage measures to reduce a sparse and high-dimensional input space to a compact representation. This compact representation is used to guide the fuzzer towards unexplored regions of the source code. Further, this compact representation can be transferred across programs that operate on the same input format. Our findings suggest \tool can improve edge coverage significantly while discovering several previously unseen bugs.

\section*{Acknowledgements}
This work is sponsored in part by NSF grants CNS-18-
42456, CNS-18-01426, CNS-16-17670, CNS-16-18771,
CCF-16-19123, CCF-18-22965, CNS-19-46068, CCF 1845893, CNS 1842456, and CCF 1822965. This work is also sponsored by ONR grant
N00014-17-1-2010; an ARL Young Investigator (YIP) award;
a NSF CAREER award; a Google Faculty Fellowship; a
Capital One Research Grant; and a J.P. Morgan Faculty Award.
Any opinions, findings, conclusions, or recommendations expressed herein are those of the authors, and do not necessarily
reflect those of the US Government, ONR, ARL, NSF, Google,
Capital One or J.P. Morgan. 
